\documentclass[12pt,a4paper]{article}

\usepackage{amsmath, cite}
\usepackage{cancel}
\usepackage{amssymb}
\usepackage{mathrsfs}
\usepackage{graphicx,subcaption}
\usepackage{dsfont}
\usepackage[colorlinks=true,linkcolor=black,citecolor=black,urlcolor=black]{hyperref}
\usepackage{enumerate}
\usepackage{epstopdf}

\newcommand{\bea}{\begin{eqnarray}}
\newcommand{\eea}{\end{eqnarray}}
\newcommand{\be}{\begin{equation}}
\newcommand{\ee}{\end{equation}}

\def\P{\mathscr P}
\def\Z{\mathds Z}

\def\O{\mathcal O}

\newdimen\hpt
\newdimen\npt
\newdimen\nhpt
\def\boxslb#1#2#3{\raisebox{8\hpt}{\raisebox{#1\nhpt}{\text{$\hspace{10pt}\underset{\text{\small{#3}}}{\framebox(#1,#1){\hspace{-10pt}\hspace{#1\npt}#2}}$}}}}

\newcommand{\Dslash}{\ensuremath \raisebox{0.025cm}{\slash}\hspace{-0.32cm} D}
\newcommand{\C}[1]{$(\ref{#1})$}

 \numberwithin{equation}{section}

\begin{document}
%%%%%%%%%%%%%%%%%%%%%%%%%%%%%%%%%%%%%%%%%%%%%%%%%%%%%%%%%%%%%%%%%

\title{\begin{flushright}\vspace{-1in}
       \mbox{\normalsize EFI-15-35}
       \end{flushright}
       \vskip 20pt
Global Anomalies and Effective Field Theory }

\date{\today}

\author{
Siavash Golkar$^{a,b}$\thanks{\href{mailto:golkar@uchicago.edu}{golkar@uchicago.edu}}~
 and
 Savdeep Sethi$^{b}$\thanks{\href{mailto:sethi@uchicago.edu}{sethi@uchicago.edu}}
%   Dam T. Son
%   \thanks{\href{mailto:dtson@uchicago.edu}
%     {dtson@uchicago.edu}}
%      \\ \\
%   {\it \it Kadanoff Center for Theoretical Physics and Enrico Fermi Institute,}\\
%   {\it   University of Chicago, 5640 South Ellis Ave., Chicago, IL 60637 USA}
%} 
 \\ \\
%\vskip 0.2 cm
$^{a}${\it Kadanoff Center for Theoretical Physics, University of Chicago}\\
$^{b}${\it Enrico Fermi Institute, University of Chicago, Chicago, IL 60637, USA}}

\maketitle\thispagestyle{empty}

\begin{abstract}

We show that matching anomalies under large gauge transformations and large diffeomorphisms can explain the appearance and non-renormalization of  couplings in effective field theory. We focus on thermal effective field theory, where we argue that the appearance of certain unusual Chern-Simons couplings is a consequence of global anomalies. As an example, we show that a mixed global anomaly in four dimensions fixes the chiral vortical effect coefficient. This is an experimentally measurable prediction from a global anomaly. For certain situations, we propose a simpler method for calculating global anomalies which uses correlation functions rather than eta invariants.

%	We show that considerations of anomalies under large gauge and diffeomorphism transformations can have non-trivial implications for the infrared theory. We focus in particular on thermal partition functions and demonstrate that the effects of global anomalies manifests as an unusual Chern-Simons term. As an example, we show that there is a mixed global anomaly in certain backgrounds in 4-dimensions and demonstrate that this anomaly fixes the chiral vortical effect coefficient. As far as we know, this is the first experimentally measurable prediciton of global anomalies. Under certain conditions, we propose a simpler method of calculating global anomalies using correlation functions.
\end{abstract}

\newpage

\setcounter{page}{1}\section{Introduction}
\label{sec:intro}
%%%%%%%%%%%%%%%%%%%%%%%%%%%%%%%%%%%%%%%%%%%%%%%%%%%%%%%%%%%%%%%%%

Among the tools at the disposal of a quantum field theorist, anomaly matching is one of the most powerful~\cite{tHooft:1979bh}. This is the idea that an anomaly can be computed at any scale; for example, using either ultraviolet or infrared degrees of freedom.  It is particularly useful as a way to determine specific couplings that might be required in an effective field theory to reproduce an anomaly. In recent years, this tool has been used to study the low energy behavior of thermal partition functions for systems with chiral anomalies. The anomalous ward identities associated to these continuous symmetries require specific Chern-Simons terms in the effective action. In turn, these couplings  have both theoretical applications as well as experimental consequences for various measurable response functions \cite{Son:2009,Neiman:2010,Landsteiner:2011cp,Landsteiner:2011iq,Banerjee:2012,Jensen:2012-1,Jensen:2013,Komargodski:2014}.

%The idea of 't Hooft anomaly matching is one of the most powerful tools in the possession of field theorists. It has countless uses in the form of restricting effective field theories as well as providing non-trivial checks for various dualities. In recent years, this tool was applied to analyze the low energy behavior of thermal partition functions of 4-dimensional systems with a chiral anomaly. It was seen that the the anomalous ward identities associated to these continuous symmetries manifests as Chern-Simons terms in the effective action. These terms in turn have experimental consequences for various response functions \cite{Son:2009,Neiman:2010,Banerjee:2012,Jensen:2012-1,Jensen:2013}. 

Most of the discussion of anomaly matching involves symmetry transformations which are continuously deformable to the identity, or trivial transformation. The associated anomalies are usually called perturbative or infinitesimal anomalies. In this work, we will be concerned with global anomalies, which involve symmetry transformations that cannot be continuously deformed to the identity; for example, either large gauge transformations or large diffeomorphisms. The associated anomalies are sometimes called global or non-perturbative anomalies. 

The existence of a global anomaly can also require that specific couplings be present in an effective field theory description. To date, however, there has been little discussion about how global anomalies can be used to predict the coefficients of couplings in an effective action. This is, in part, because of added complications that arise when dealing with global rather than perturbative anomalies. For example, in many theories with a symmetry group that includes large transformations, there are already anomalies under infinitesimal symmetry transformations. This makes it difficult to write down sensible anomalous ward identities for just global transformations. A more practical limitation is that in experimental setups, it is difficult to construct interesting topologies that would allow one to directly probe physical phenomena connected to large gauge transformations.

%Similar to anomalies of continuous symmetries, the presence of global anomalies\footnote{In this paper we denote anomalies that are not continuously deformable to the identity as large, global or non-perturbative anomalies. In the case of gravitational anomalies these can also be referred to as modular anomalies. On the other hand, anomalies that are connected to the identity will be referred to as perturbative or infinitesimal anomalies.} amounts to restrictions on field content of effective theories as well as on dualities between different theories.  However, to this date there have been no direct consequences of global anomalies in the direction of predicting coefficients of effective actions. This is due to the existence of various complications when dealing with global anomalies. For example, in many theories with anomalies under large transformations of some group, there also exists an anomaly under infinitesimal transformations of the same group. This makes it difficult to write down sensible anomalous ward identities for the global part. A more practical limitation is that in experimental setups, there is usually no interesting topology that would allow for large gauge transformations.

Despite these difficulties, one of us recently conjectured that the coefficient of a particular transport phenomenon, known as the chiral vortical effect, is related to the presence of a global anomaly~\cite{Golkar:2012}. This conjecture is part of the motivation for this work. In a broader context, consider a quantum field theory compactified on a Euclidean space-time of even dimension $d$ with a circle fibration. Let $t$ denote the circle coordinate and $x^i$ denote the remaining coordinates. The metric for such a space-time takes the form,
\be
ds^2 =e^{2\sigma(x)}\left(dt+a_i(x) dx^i \right)^2 + h_{ij}(x)dx^i dx^j.
\ee
The vector-field $a_i$ is the background graviphoton. As an illustration, assume that the field theory contains an abelian gauge-field $A$ with field strength $F=dA$. Under suitable restrictions on the space-time, one can find a Chern-Simons coupling in the $d-1$-dimensional effective field theory, obtained by integrating out the $t$ direction, of the form,
\be\label{generalCS}
\int a\wedge F \wedge \cdots \wedge F, 
\ee 
where $a=a_idx^i$ is the graviphoton $1$-form. %Usually, there are also couplings involving various combinations of metric curvatures and gauge field strengths. 
A coupling like~\C{generalCS}\ is peculiar because it involves a naked graviphoton rather than a momentum suppressed interaction involving the spin connection, which one might expect. Explaining why this coupling appears and how it is connected with global anomalies is a basic goal of this work. Although this example involves a circle-fibered space-time, there is a similar story for toroidally-fibered space-times, which are particularly useful when studying field theories in odd dimensions. For example, the relationship between the chiral vortical effect coefficient (CVE) and global anomalies, conjectured in~\cite{Golkar:2012}, involves compactification on a $4$-torus.  This is a concrete case where the presence of a global anomaly has experimentally measurable consequences.

A corollary of the appearance of global anomalies in local effective actions is that we can compute the change of the action under a large gauge transformation directly via correlation functions of local operators. In general, this is simpler than the original method of computation via an $\eta$ invariant~\cite{Witten:GGA}. There are other advantages of this approach. For anomalies that are computable this way, reciprocity of mixed global anomalies becomes manifest. This is the phenomenon that, in certain cases,  the presence of a gravitational anomaly in a gauge background implies the presence of a gauge anomaly in a gravitational background.
%As we will see, some properties of global anomalies that are more opaque in other approaches %less transparent from  \cite{Witten:GGA}  
%become very clear in our analysis. One example is the reciprocity property of mixed global anomalies where the presence of a gravitational anomaly in a gauge background necessitates the presence of a gauge anomaly in a gravitational background. 
This result is clear for perturbative anomalies using the descent formalism from an anomaly polynomial, where gauge and gravitational curvatures appear on equal footing. %However, the role of the mapping torus in the derivation of global anomalies ruins the symmetry between the gauge and gravity sectors, and reciprocity is not as clear. 
When such reciprocity exists for global anomalies, it becomes manifest in the effective action approach.

%In many ways in fact we can argue that we can interpret our results as a local manifestation of the $\eta$ invariant governing global anomalies.  

The paper is organized as follows: in section~\ref{sec:GAM}, we give an overview of our methodology and go over the main points of our analysis. We also give a quick review of the original argument by Witten relating global anomalies to $\eta$ invariants~\cite{Witten:GGA}. In section~\ref{sec:examples}, we work through a number of examples in 2, 3 and 4 dimensions. The 2-dimensional case is worked out in considerable detail since it is central to our other examples. We also comment on how this case can have simple generalizations in higher dimensions.

%%%%%%%%%%%%%%%%%%%%%%%%%%%%%%%%%%%%%%%%%%%%%%%%%%%%%%%%%%%%%%%%%
 \section{Global Anomaly Matching}
 \label{sec:GAM}
%%%%%%%%%%%%%%%%%%%%%%%%%%%%%%%%%%%%%%%%%%%%%%%%%%%%%%%%%%%%%%%%%

Our argument relies on two facts: first that  the partition function of a theory without gapless excitations must be a local functional of the background fields; therefore, the effective action is local. Second that this effective action must reproduce the anomalies of the microscopic theory. We will later discuss some methods of consistently gapping an anomalous theory, and discuss the implications of global anomalies for the effective action describing the IR theory. Since the consequences of perturbative anomalies are well studied, we will henceforth assume that either perturbative anomalies are absent, or that they have been matched via other terms in the effective action.

%As we stated in the introduction, our argument relies on two facts. First,  the partition function of a theory with no gapless excitations must be a local functional of the background fields therefore the effective action is local. And second, this effective action must reproduce the anomalies of the microscopic theory. In what follows, we will discuss methods of consistently gapping an anomalous theory and discussing implications of global anomalies in the effective action describing the IR. Since the consequences of the effects of infinitesimal anomalies is well studied, we will henceforth assume that either infinitesimal anomalies are not present or they have been matched via other terms in the effective action.

It is important to note that, essentially by definition, the existence of an anomaly prevents the generation of a gap using interactions that preserve all the symmetries of the theory. This follows from the preceding argument because the low-energy effective action would have to both reproduce the anomaly and be simultaneously local. However, we know that there is no local counter-term that can remove the effect of the anomaly, while simultaneously preserving all other symmetries. This proves that gapping the spectrum of an anomalous theory requires some breaking of symmetries. 

%It is important to note that, essentially by definition, the existence of an anomaly prevents the generation of a gap via interactions that preserve all the symmetries of the theory. This is due to the fact that using the above argument the effective action would have to reproduce the anomaly and be local at the same time. But we know that there is no local counter-term that can be written down that can remove the effects of the anomaly and at the same time preserves all the other symmetries. This proves that gapping the spectrum of an anomalous theory requires some breaking of symmetries. 

In this work, we primarily consider the thermal partition function of fermions in space-time dimension $d$. In the high temperature, or equivalently low energy limit, the thermal circle cannot be resolved and the theory is effectively defined on $d-1$ dimensions. The anti-periodic boundary conditions imposed on the fermions produces a mass gap proportional to the temperature so from an effective field theory (EFT) perspective, the theory is gapped. Hence the effective action must be a local functional of the background fields. In cases where the microscopic theory is anomalous, the EFT must reproduce the anomaly. However, since  the gap is generated by breaking the $d$-dimensional rotational symmetry, we can circumvent the inability of a local counter-term to reproduce the anomaly. We do not claim that this is the only mechanism that can generate a gap, but any other mechanism must break some symmetry in a similar fashion.

%In this paper, we primarily deal with the case of the thermal partition function of fermions defined in various space-time dimensions $d$. In the high temperature limit (equivalently low energy limit), the thermal circle cannot be resolved and the theory is effectively defined on $d-1$ dimensions. The anti-periodic boundary condition imposed on the fermions produces a mass gap proportional to the temperature and from the EFT perspective, the theory is gapped. Hence, the effective action must be a local functional of the background fields. In cases where the microscopic theory is anomalous, the EFT must reproduce this anomaly. However, since in the process of generating a gap we broke the $d$ dimensional rotational symmetry, we can circumvent the no anomaly counter-term issue mentioned above. We do not claim that this is the only mechanism that can generate a gap but any other mechanism must break some symmetry in a similar fashion.

In the remainder of this section we describe some of the restrictions and features of our approach.

\subsubsection*{I. Global symmetry transformations must be compatible with the EFT}

Since we are interested in matching anomalies using local functionals, it is crucial that we analyze the system below the scale of the gap. For example, if the gap is generated by thermal boundary conditions, we can only study background fields that carry energy less than the temperature. 

It is therefore crucial that the global transformations we consider and the reduction to an EFT be mutually compatible.  For example in 2 dimensions, the group of large diffeomorphisms of a torus is the modular group $SL(2,\mathds Z)$ generated by transformations $T$ and $S$. These generators act by sending, 
\be T: (x, t) \to (x, t+x), \qquad S: (x, t) \to (t, -x), \ee 
respectively. We have taken canonical periodicities for the torus coordinates:
\be
(x,t) \sim (x,t) + 2\pi (n,m) \qquad n,m\in \Z. 
\ee 
Consider a Weyl fermion with thermal boundary conditions in the time direction.  We want to reduce along $t$ to get a $1$-dimensional effective field theory. The only large diffeomorphisms compatible with this reduction are transformations which are $t$-independent. Such transformations preserve the condition that the background data carry energy small compared to the scale of the gap.  It is easy to see that this condition is respected by the $T$ transformation. We therefore expect to capture an anomaly under the $T$ transformation by a local functional. However since $S$ exchanges $x$ and $t$, it does not preserve the region of validity of the EFT; hence there is no reason to expect that an anomaly under $S$ would be captured by a local functional of the metric.

\subsubsection*{II. Certain correlators compute global anomalies}

Once we know that the global anomaly must be matched using a functional of the background fields, we can determine various coefficients in the effective action through anomaly matching. For example, in the next section we will see that the chiral vortical effect  coefficient can be fixed once we  determine the 4-dimensional anomaly.

However, we can also reverse the argument and use the preceding line of reasoning to predict and compute certain global anomalies via correlation functions. This computation follows the standard EFT procedure. We write down all possible terms which preserve the symmetries of the IR theory, do not produce any perturbative anomalies, and yet are not invariant under global transformations. An example of such term is $\int a \wedge dA$, which is a special case of~\C{generalCS}. The requirement of being invariant under perturbative but not global transformations is very restrictive in general. Typically, there are only a few terms that can produce global anomalies. We can therefore compute the anomaly by calculating the related correlation function; for example, in the case of $\int a \wedge dA$ we would calculate the stress-tensor/current correlator $\langle T^{tx}J^y \rangle$. This is the same correlator that is responsible for the finite temperature CVE \cite{Amado:2011zx}. We stress again that the only global anomalies computable this way are those generated by transformations compatible with the EFT reduction.

\subsubsection*{III. Global anomalies and the $\eta$ invariant}

In his original discussion, Witten related the global gravitational anomaly to the $\eta$ invariant of a mapping torus constructed from the transformation under consideration~\cite{Witten:GGA}. Here we give a brief review of this method keeping in mind that unlike~\cite{Witten:GGA}, many of the cases with which we are concerned also possess a perturbative anomaly. This will lead to situations where the $\eta$ invariant is not purely a topological number. Indeed, in order to extract a meaningful number, we need to subtract the contribution of an associated Chern-Simons term, which can be interpreted as removing the perturbative anomaly via a Green-Schwarz mechanism.

%It is known that global anomalies are related to the $\eta$ invariant of a mapping torus constructed from the transformation \cite{Witten:GGA}. Here, we give a brief review of the method keeping in mind that the unlike in \cite{Witten:GGA} the cases we are concerned with also possess a vituperative anomaly. This will lead to the property that the $\eta$ invariant is not purely a topological number. Indeed in order to extract a meaningful number we need to subtract the contribution of an associated Chern-Simons term which can be interpreted as removing the anomaly via a Green-Schwarz mechanism.

The general method goes as follows: take a compact even-dimensional ($d=2n$) manifold $\mathcal M$ endowed with a metric $g_{\mu\nu}$ and a possible gauge-field  $A_\mu$. We are interested in the change of the partition function under a symmetry transformation $\pi: g_{\mu\nu} \to g_{\mu\nu} ^{(\pi)}, \; \; A_{\mu} \to A_{\mu} ^{(\pi)}$. To calculate this change, we construct interpolating metrics and gauge-fields
\begin{equation}
\label{eq:Interpolating_function}
g_{\mu\nu}(y) = (1-y) \,g_{\mu\nu} + y \, g_{\mu\nu} ^{(\pi)},\;\;\;
A_{\mu}(y) = (1-y) A_{\mu} + y \, A_{\mu} ^{(\pi)},
\end{equation}
which go smoothly between the background fields and their transformations under $\pi$. We then construct a higher dimensional manifold by promoting the parameter $y$ to a coordinate with metric:
\begin{equation}
\label{eq:Mapping_Torus}
ds^2 = dy^2 + g_{\mu\nu}(y) dx^\mu dx^\nu.
\end{equation}
We also trivially extend the gauge field into the bulk. The manifold at $y=0$ is identified with the manifold at $y=1$ resulting in a compact space called the mapping torus for the transformation $\pi$. We will denote the mapping torus by $\Sigma$. Let us specialize to the case of a Weyl fermion $\psi$. The change of the partition function under the transformation $\pi$ is given by a phase $\eta$:
\begin{equation}
\label{eq:Action_Change}
Z(g_{\mu\nu},A_\mu) \to Z(g_{\mu\nu}^{(\pi)},A_\mu^{(\pi)})=e^{i\pi \eta} Z(g_{\mu\nu},A_\mu).
\end{equation}
The $\eta$ parameter is defined as a regulated sum of the signs of the eigenvalues of the Dirac operator defined on $\Sigma$, 
\begin{equation}\
\Dslash \psi= \lambda \psi, \qquad \eta=\sum_\lambda \text{sign }(\lambda).
\end{equation}
The direct computation of the $\eta$-invariant is possible, but can be challenging because of the boundary conditions imposed on the mapping torus along the $y$ circle. However, there is a simpler way to compute $\eta$ using index theory. Specifically, by studying the  Atiyah-Patodi-Singer (APS) index theorem applied to the spin complex \cite{EGH}.  Assume that the mapping torus $\Sigma$ is the boundary of some $d+2$ dimensional manifold $X$, and that the Dirac operator extends to an operator $\Dslash_X$ on $X$. The index of the Dirac operator on $X$ with APS boundary conditions is given by: 
\begin{equation}
\label{eq:APS}
{\rm Ind}(\, \Dslash_X )= \int_X \widehat{A}(X)\wedge \text{ch}(V)+\int_{\Sigma} CS-\frac12 \eta. 
\end{equation}
where $\int CS$ denotes a Chern-Simons term that removes the dependence of the right hand side on the choice of embedding of the boundary. Note that the $\eta$ invariant defined by~\C{eq:Action_Change}\ determines the phase of the partition function under both small and large symmetry transformations. If a perturbative anomaly is present in the fermionic path integral, but is removed from the theory using a Green-Schwarz mechanism, one would have to subtract a Chern-Simons contribution from $\eta$ to find the topological number that purely reflects the effect of the global transformation.\footnote{In other words, the $\eta$ invariant is not topological but the difference of the $\eta$ invariant and the Chern-Simons term is indeed topological since it is invariant under small deformations of the metric and gauge field.} We give an example of this calculation in appendix~\ref{app:2D_mappingtorus}. 

Note that the details of the computation of the global anomaly are not crucial for our discussion. In each case we consider, the theory can be dimensionally reduced to 2 dimensions, where the effects of global transformations are well known. It is worth iterating that the $\eta$ invariant appearing in the index formula is related to topological data of a higher-dimensional auxiliary manifold. However in the cases we study, the $\eta$ invariant will also appear as the coefficient of a local functional in an effective action, computable from correlation functions.

\subsubsection*{IV. Reciprocity of mixed anomalies}

A consequence of descent relations for perturbative anomalies is that if there are terms in the anomaly polynomial which mix gauge and gravitational curvatures, there will be mixed perturbative anomalies: a gravitational anomaly in the presence of a gauge background as well as a gauge anomaly in the presence of a gravitational background. Consistency would then require that one cannot exist without the other. We call this property reciprocity of mixed anomalies.

Since global anomalies are also derived from the anomaly polynomial using index theory, one might expect that a similar consistency condition should also  hold for mixed global anomalies. However, this analysis is more complicated for two reasons. First, the presence of large transformations depends on the structure of the group of gauge transformations and diffeomorphisms. For example, we might consider a space that admits large diffeomorphisms but a gauge group that admits no large gauge transformations. There is no possible reciprocity of global anomalies in such a case.  Second, the derivation of the global anomaly from the mapping torus breaks the symmetry between the two sectors. 

However, in special cases where both sectors allow large transformations, one can see that a reciprocity property is required by consistency. We will encounter one such example in section \ref{sec:4D}. As we will see, from the effective field theory perspective, this reciprocity property is manifest, since the change of the partition function comes from the variation of a single term in the effective action.

%In particular, under certain circumstances For example, let us calculate the global gravitational anomaly in $d$ dimensions in the presence of some gauge flux. In order to do so, we construct a mapping torus which interpolates between the manifold and its transform as a function of a higher dimensional coordinate $y$. The resulting mapping torus is a manifold which has both gauge and gravitational fluxes. In particular, it can be also interpreted as a mapping torus for a large gauge transformation in the presence of a gravitational flux. \note{Is this really true in general?} This proves our assertion; for an example, see appendix~\ref{app:reciprocity}. It is therefore crucial that after gapping the theory, the local effective action must have this ``reciprocity'' property as well. As we will later see, this will indeed be the case.

\subsubsection*{V. Decompactification limit}

In the derivation of the effective action, we have assumed that the spatial manifold is compact with a non-trivial modular group. A simple example would be a 3-torus denoted $T^3$. However, thermal partition functions of interest are usually defined on $\mathds R^3 \times S^1$. Therefore, in order to make contact with physics in the real world, we should decompactify the $T^3$ and show that our conclusions about the effective action survive. A short argument shows that this is indeed the case. Take the size of the thermal circle to be $\beta$ and the size of the spatial manifold to be set by $L$. We are interested in the limit that the wavelength of perturbations $\lambda$ satisfies $\beta<<\lambda<<L$. 

Now we analyze the problem via a Wilsonian renormalization perspective. In order to derive the effective action, we have integrated out short length scale degrees of freedom that are much smaller than $\lambda$. These short length scale degrees of freedom do not carry any information about the large scale properties of the system. In particular they do  not know about the topology of the spatial manifold. They do, however, know about the boundary conditions of the temporal circle which is small. This argument implies that if we scale up the size of the spatial manifold, we would find the same effective action; hence the decompactification limit is continuous. We show explicitly that this continuity is true for the special cases of 2 and 4 dimensions in the following section. %An extension of this argument would tell us that we can even replace $T^3$ with $S^3$ or any other compact manifold and the effective action would be the same. 

%%%%%%%%%%%%%%%%%%%%%%%%%%%%%%%%%%%%%%%%%%%%%%%%%%%%%%%%%%%%%%%%%
 \section{Examples}
 \label{sec:examples}
%%%%%%%%%%%%%%%%%%%%%%%%%%%%%%%%%%%%%%%%%%%%%%%%%%%%%%%%%%%%%%%%%

In this section we give some examples to illustrate the various points of our approach. We start with the case of Weyl fermions in two dimensions and generalize to higher-dimensional cases. Almost all our higher-dimensional examples can be derived from the two-dimensional case via dimensional reduction and the use of index theorems. 
 
%%%%%%%%%%%%%%%%%%%%%%%%%%%%%%%%%%%%%%%%%%%%%%%%%%%%%%%%%%%%%%%%%
\subsection{Weyl Fermions in 2D}
\label{sec:2D}
%%%%%%%%%%%%%%%%%%%%%%%%%%%%%%%%%%%%%%%%%%%%%%%%%%%%%%%%%%%%%%%%%
The anomaly properties of Weyl fermions in 2D are well studied. In particular the modular properties on a torus are well known. Since there are no mixed anomalies in 2 dimensions, we will only concern ourselves with global gravitational anomalies and set any gauge fields to zero. The case of a pure gauge anomaly would be discussed similarly.

We want to analyze the consequences of global anomalies in the low energy limit after gapping the system. As discussed in the previous section, one way of introducing a gap consistently is to look at the thermal partition function, i.e. define the system on $\mathds R \times S^1$ and impose anti-periodic boundary conditions along the circle direction. This would ensure that there are no zero modes in the system and the lowest lying excitation has energy proportional to the inverse size of the circle (i.e., the temperature $\beta^{-1}$ for a thermal partition function). Therefore, at energy scales below the gap, the effective action will be local. 
In the following discussion, we will first compactify the spatial direction and consider the torus partition function. At the end, we will take the decompactification limit. 
%In what follows, we also compactify the spatial direction and consider the partition function on a torus and take the decompactification limit in the end.

To be concrete, define the system on a torus parametrized by $(t,x)$ with metric:
\begin{equation}\label{eq:2d_constant_metric}
ds^2=e^{2\sigma(x)}\left(dt+a(x) dx\right)^2 + dx^2.
\end{equation}  
We impose anti-periodic boundary conditions along the $t$ circle.\footnote{Strictly speaking, in this case anti-periodic boundary conditions are not required to arrive at a local action that reproduces the global anomaly. We will comment on this later in the section.} We take the $t$ circle to have periodicity $\beta$ and $x$ to have periodicity $L$. It is important to note that all the background fields are $t$-independent. Indeed any $t$-dependence would introduce energies of the scale $\beta^{-1}$ for which we have no reason to expect the theory to be local. 

The group of large diffeomorphisms of the torus is $SL(2,\mathds Z)$ generated by transformations 
\be T: (x, t) \to \left(x, t+{\beta x \over L}\right), \qquad S: (x, t) \to (t, -x). \ee 
%$T: t\to t+{\beta x \over L}, \,x\to x$ and $S:t\to x,\, x\to-t$. 
Of these two transformations, only the $T$ transformation is compatible with the form of our metric \eqref{eq:2d_constant_metric} and the transformation which sends $a(x) \to a(x) +{\beta \over L}$. The $S$ transformation would introduce high energy modes and would drive the system out of the regime of validity of the local effective action. 

Therefore the goal of this section is first to derive the transformation properties of the partition function under the $T$ transformation, and then to derive a local effective action that can reproduce this transformation in the low energy limit. Before proceeding, we will give a short review of  perturbative and global anomalies.

\subsubsection*{Anomalies}\label{subsec:local_2D_anomalies}
Before considering global properties of the system, let us review the perturbative anomalies. We consider a theory of Weyl fermions with different chiralities defined on a space with metric $g_{\mu\nu}$. Such a theory has a gravitational anomaly that is given via descent equations from the 4-dimensional anomaly polynomial:
\begin{equation}
\mathcal P_{anom}[F,R]= c_g\, \text{tr} (R\wedge R).
\end{equation}
The coefficient $c_g$ is the pure gravitational anomaly given by,
\begin{align}
c_g=&-\frac{1}{96\pi}\sum_{i} \chi_i, 
\end{align}
where $\chi_i = \pm 1$ denotes the chirality of the particles. We take the convention that right-handed fermions have positive chirality. The contribution of a Majorana-Weyl fermion would be ${1\over 2}$ of a Weyl fermion in this sum. If the theory is conformal, this sum simplifies
\begin{align}
c_g=&-\frac{1}{96\pi}(c_R-c_L), 
\end{align}
where $(c_R, c_L)$ are the central charges of the right and left sectors, respectively. The anomaly polynomial is the exterior derivative of $3$-dimensional Chern-Simons couplings, 
\begin{equation}
\label{eq:CS3}
I_{CS}= c_g \int ( \omega\wedge  d\omega +\frac23 \omega\wedge\omega\wedge\omega).
\end{equation}
A gauge variation of $I_{CS}$ on a 3-manifold is necessarily a total derivative. If the 3-manifold has a boundary supporting our 2-dimensional fields then the gauge variation of~\C{eq:CS3}\ gives the anomalous variation. 
%If this term is defined on 3-Manifolds with 2-dimensional boundary, its variation under a diffeomorphism transformation gives the anomaly of the two dimensional boundary theory.  However, the exact form of the perturbative anomaly is not important for us. 

The $\eta$ invariant computed on the mapping torus associated to the gauge transformation under consideration has a topological contribution, corresponding to any global anomaly, as well as a Chern-Simons contribution, corresponding precisely to the local anomaly. 
%As mentioned in the previous section, the transformation is computed by the $\eta$ invariant which has a topological piece as well as the Chern-Simons contribution which precisely gives the local anomaly. 
Since we are interested in the global anomaly contribution, we will need to subtract this Chern-Simons contribution from the $\eta$ parameter calculation. An explicit derivation of the global anomaly from the mapping torus is provided in appendix~\ref{app:2D_mappingtorus}.

\subsubsection*{Modular properties on a torus}
 
We wish to analyze the transformation properties of our system under the $T$ transformation, which sends $(x,t) \to \left(x, t+{\beta x \over L}\right)$. For simplicity, we take the metric~\C{eq:2d_constant_metric} with $\sigma =0$,
\begin{equation}
ds^2=(dt+a(x) dx)^2 + dx^2.
\end{equation}  
We have a choice of spin structure along both the $t$ and $x$ directions. Let us denote this choice with a square whose vertical axis is $t$ and whose horizontal axis is $x$.  For example $\boxslb{12}{A}{P}$ denotes the partition function of the system with anti-periodic boundary condition in the $t$ direction and periodic boundary condition along the $x$ direction.\footnote{This notation is similar to standard notation in string theory; see, for example~\cite{Ginsparg}, except we have flipped the roles of $x$ and $t$ in anticipation of the generalization to higher dimensions. }
%It is worth noting that since in the free case the theory is conformal and our background is flat, we are allowed to do any rescaling we wish.

The $T$ transformation of the various spin structures are well-known~\cite{Ginsparg}, with a direct calculation of these phases provided in appendix~\ref{app:2D_mappingtorus}:
\begin{align}
\boxslb{18}{A}{A}\to  e^{-\tfrac{i\pi}{24}} &\, \boxslb{18}{A}{P}, \; 
\hspace{40pt}
\boxslb{18}{A}{P}\to  e^{-\tfrac{i\pi}{24}} \, \boxslb{18}{A}{A} \notag, 	\\
&\;\boxslb{18}{P}{A}\to  e^{\tfrac{i\pi}{12}} \, \boxslb{18}{P}{A}\;.  
\end{align}
%The change in the periodicity is easy to understand. The $T$ transformation is generated by the re-parametrization $t \to t'=t + \beta x / L$ and hence the periodicity under $t$ doesn't change but periodicity under $x$ sees the sign flip from both the $x$ and the $t$ direction (because $t'$ also changes with $x$). 
Note that partition function for $\boxslb{12}{P}{P}=0$ because of the presence of a fermionic zero mode. 

We want to sum over a combination of spin structures that are left invariant under the $T$ transformation. 
%In order for our discussions to make sense we must pick a transformation and spin structure combination that are invariant. 
So we can pick either $\boxslb{12}{P}{A}\,$, which corresponds to an insertion of $(-1)^F$, or $\boxslb{12}{A}{A}+\boxslb{12}{A}{P}\,$, which is a thermal partition function that sums over both periodicities in the spatial direction. We can also look at the transformation $T^2$, which leaves all boundary conditions invariant. This has the added advantage of allowing us to analyze each structure individually. Under $T^2$, 
\begin{align}
\boxslb{18}{A}{A}\to  e^{-\tfrac{i\pi}{12}} &\, \boxslb{18}{A}{A}\;, \; 
\hspace{40pt}
\boxslb{18}{A}{P}\to  e^{-\tfrac{i\pi}{12}} \, \boxslb{18}{A}{P} \;\notag,	\\
&\;\boxslb{18}{P}{A}\to  e^{\tfrac{i\pi}{6}} \, \boxslb{18}{P}{A}\;,
\end{align}
and the field $a \to a + {2\beta \over L}$.

\subsubsection*{The effective action}
We can now read off which terms are needed in the effective action in order to reproduce the $T^2$ transformation of these different sectors. We first consider the sectors which are anti-periodic in  time. We iterate that we are working at energy scales much smaller than the temperature (length scales much larger than the $t$ circle size $\beta$) and the IR theory is $1$-dimensional. The effective action must therefore be a local 1D functional of the background fields; here the only non-zero field is $a(x)$. Since the perturbative anomaly vanishes in this flat background, we need to classify all the possible terms in the effective action which are invariant under small, but not large diffeomorphisms. 

There is only one such term, which is the $1$-dimensional Chern-Simons term $\int a$. Matching the variations of the action, we see that the coefficient is fixed to be:
\begin{equation} \label{unusualCS}
S_{eff}= \beta^{-1}\frac{i\pi}{24}\int a.
\end{equation}
The presence of this term is required to match the global anomaly.\footnote{We note that this term was considered in~\cite{Jensen:2012}. By looking at the theory on manifolds with conical singularities and assuming continuity in the flat limit,  \cite{Jensen:2012}\ argued that the coefficient can be related to a perturbative gravitational anomaly. From our argument, we see that this is not always the case because it is possible to have a theory with no perturbative gravitational anomaly, but with a non-zero coefficient for $\int a$.}

The Chern-Simon's coupling~\C{unusualCS}\ is unusual because it involves a naked graviphoton. This should be contrasted with the usual gravitational Chern-Simons term, which is defined in terms of the spin connection. The spin connection is already suppressed by $1$ momentum relative to the graviphoton.   
%However, it is unusual in that it is a Chern-Simon in which a component of the metric is appearing without any derivatives. For example the usual gravitational Chern-Simons term is defined in terms of the spin connection which is a one derivative object compared to the metric.
 The appearance of this unusual term provides an alternate way of computing the global anomaly in terms of a correlation function. 
 % allows us to in fact calculate the global anomaly which is the coefficient of this unusual term as a correlation function. 
 From the metric we see that,
\begin{equation}
\label{eq:a_variation}
\delta a= \delta g_{tx}+\delta g_{xt}+2 a \delta g_{xx},
\end{equation}
where we have not imposed symmetry in the indices in taking the variation. This means we can relate the global anomaly to the expectation value of the stress tensor $\langle T^{tx} + a\, T^{xx}\rangle$. In particular for a diagonal background metric, the global anomaly under $T$ is given by
\[ \langle T^{tx}\rangle_{g_{\mu\nu}=\delta_{\mu\nu}}.\]
Instead of computing the change of the partition function under a large diffeomorphism, we can deduce the existence and value of the global anomaly for $T$ by looking at this correlation function.

Finally, we look at the sector with periodic boundary conditions along the $t$ circle. At length scales much larger than $\beta$, that is from the 1-dimensional perspective, this theory is not gapped; the fermion has a Kaluza-Klein zero mode $\psi_0$. Therefore, we do not expect our general methodology to hold in this case.  However, $\psi_0$ has zero Kaluza-Klein momentum which implies that it is not charged under the graviphoton $a$. Since we are interested in the large diffeomorphisms that take $a \to a + {\beta \over L}$, we do not expect this zero mode to contribute. Stated another way, the zero KK momentum sector of the dimensional reduction of the $2$-dimensional fermion, $\psi$, is invariant under these transformations. We therefore expect the action to factorize into a sum of two terms: one that knows about the background field $a$ and another piece with the zero winding. Again matching the global anomaly gives:
\begin{equation}
S_{eff}= -\beta^{-1}\frac{i\pi}{12}\int a + S(\psi_0).
\end{equation}
For non-vanishing partition functions, we can again compute the global anomaly in this sector using the same correlators as before.

\subsubsection*{Decompactification limit}
It is important to note that although our arguments for matching global anomalies hold for compact spaces, we can still derive non-trivial results by taking the decompactification limit and using continuity. For example, the thermal partition function of a Weyl fermion on $S^1 \times \mathds R$ can be obtained from the large spatial circle limit of the $T^2$ case discussed above. The coefficient of the Chern-Simons coupling cannot change as we smoothly  decompactify since it must reproduce the global anomaly. We therefore expect the same quantization argument survives this limit.

\begin{figure}
	\centering
	\includegraphics[width=0.3\linewidth]{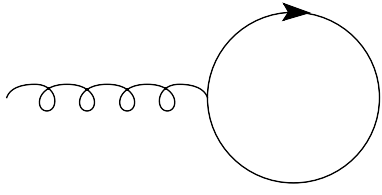}
	\caption{The 1-point stress correlator which gives the global diffeomorphism anomaly.}
	\label{fig:2d}
\end{figure}

Indeed, we can verify that this is true by a direct computation of the stress-tensor correlator on $S^1 \times \mathds R$.  We expect to find $\langle T^{01}\rangle=-\frac{i \pi}{12 }{1\over \beta^2}$ for a Weyl fermion with odd spin structure.  This is a free-field computation of the Euclidean partition function, 
\be
Z=\int D\bar\psi D\psi e^{-S}. \ee
If we put the theory on a background metric $g_{\mu\nu}=\eta_{\mu\nu}+h_{\mu\nu}$ and expand the action up to first order in $h_{\mu\nu}$ we have:
\be
S=\int \! dt\, dx\; \big(i \bar\psi \cancel \partial\psi -\frac12 h_{\mu\nu}T^{\mu\nu} +\O(h^2)\big)=\int \! dt\, dx\; \big(i \bar\psi \cancel \partial\psi - h_{01}T^{01} +\ldots\big).
\ee 
%The stress tensor is given by:
%\[T^{01}=\frac i4 \bar{\psi} \big(\gamma^0 \stackrel{\leftrightarrow}{\partial ^1}
%+\gamma^1 \stackrel{\leftrightarrow}{\partial ^0} \big)\psi.\]
%where in the second equality we have used the symmetry properties of the metric.
We are interested in evaluating:
\[\frac{\delta Z}{\delta a}\Big|_{a=0}=-\frac{\delta Z}{\delta h_{01}}\Big|_{h_{01}=0}=\langle T^{01} \rangle.\]
%To be precise, we have gamma matrices $\{\gamma^a,\gamma^b\}= 2\delta^{ab}$ and we pick the representation:
%\[\gamma^0=\sigma^1=\left( 
%\begin{split}
%0\;\;\;&1\\1\;\;\;&0
%\right), \quad \gamma^1=-\sigma^2=\left( 
%\begin{split}
%0\;\;\;&i\\-i\;\;\;&0
%\end{split}
%\right), \quad \gamma^5=\left( 
%\begin{split}
%1\;\;\;\;\;\;&0\\0\;\;\;\;\;\;&\!\!\!\!-\!1
%\end{split}
%\right) . \]
%Defining $z=x+it$, we see that the field $\psi$ can be decomposed into into holomorphic and antiholomorphic components:
%\[\psi=\left( 
%\begin{split}
%\psi_L(z)\\\psi_R(\bar z)
%\end{split}
%\right) , \]
%where we have defined left and right movers as holomorphic and antiholomorphic components respectively. 
%Hence, in our notation, $\frac12 (1-\gamma^5)$ is the projector on the homomorphic component.
As usual, $\psi$ can be decomposed into holomorphic and anti-holomorphic pieces with respect to  $z=x+it$. The diagram in question is shown in figure~\ref{fig:2d}. To calculate, we first mode expand:
\begin{equation}
\psi(t,x)=e^{2\pi i n {t \over \beta}} \psi_n(x), \qquad n \in \mathds Z+\frac12.
\end{equation}
In terms of gamma matrices $\{\gamma^a,\gamma^b\}= 2\delta^{ab}$, the propagator becomes 
\be \dfrac{1}{\cancel p}=\dfrac{p_x \gamma^1+2\pi {n \over\beta} \gamma^0}{ \left({p_x}\right)^2+4\pi^2 {n^2\over \beta^2}} .\ee  
The stress-tensor insertion becomes $T^{01}=(\frac{1}{2}\gamma^0 p_x+{n \over\beta} \pi \gamma^1)\dfrac{1-\gamma^5}2$ for a single holomorphic component. Hence the contribution of each mode is given by, 
\bea
\Big\langle\dfrac{1-\gamma^5}2 T_n^{01} \Big\rangle &=& -\int \frac{dp_x}{2\pi} \text{Tr}\left[\dfrac{p_x \gamma^1+2\pi (n/\beta) \gamma^0}{{p_x}^2+4\pi^2 (n^2/\beta^2)}\left(\frac{1}{2}\gamma^0 p_x+{n\over\beta}\pi  \gamma^1\right)\dfrac{1-\gamma^5}2\right], \cr
&=&-2 i\pi {n^2 \over \beta^2} \int dp \frac{1}{{p}^2+4\pi^2 (n^2/\beta^2)}= -i \pi { |n| \over \beta},
\eea
where we used the relation from dimensional regularization relating
\[\int \frac{d^d l}{(2\pi)^d} \frac{l^2}{l^2+\Delta}=-\Delta \int \frac{d^d l}{(2\pi)^d} \frac{1}{l^2+\Delta}.\]
We now sum the contributions from all the modes,
\be {1\over \beta} \sum_{n\in \mathds{Z}+\frac12}-i \pi {|n|\over\beta}=- {2i\pi\over\beta^2} \sum_{n=\frac12}n=-\frac{i\pi}{12\beta^2},\ee
which confirms our expectation.

%%%%%%%%%%%%%%%%%%%%%%%%%%%%%%%%%%%%%%%%%%%%%%%%%%%%%%%%%%%%%%%%%
\subsection{Dirac fermions in 3D}
\label{sec:3D}
%%%%%%%%%%%%%%%%%%%%%%%%%%%%%%%%%%%%%%%%%%%%%%%%%%%%%%%%%%%%%%%%%
In this section we analyze the case of Dirac fermions on $3$-dimensional manifolds constructed from a $T^3$, which can include non-orientable spaces. We will closely follow the recent discussion of~\cite{Ryu:2015}. 
 %on non-orientable 3-dimensional manifolds. As an example we look at the case of a flat 3-torus that was recently discussed in \cite{Ryu:2015}. 
 From $T^3$, we can construct non-orientable spaces by using twisted boundary conditions. A $T^3$ has $6$ metric moduli. We  parametrize the moduli by $3$ radii $(R_x, R_y, R_t)$ and $3$ angles $(\alpha, \beta, \gamma)$, which are angles between the $x-t$, $x-y$ and $y-t$ directions, respectively. The 3-torus has an $SL(3,\mathds Z)$ modular group. We will shortly quotient by a parity operation $\P: \;y\to -y$ to produce spaces like $S^1\times 
% We are interested in boundary conditions that involve twisting by the parity operation $\P: \;y\to -y$. In this way we construct a non-orientable manifold $S^1\times 
K$, where $K$ is the Klein bottle. The parameters $\beta$ and $\gamma$ are odd under this action so we will set them to zero. The $T^3$ metric then takes the simpler form, 
\begin{equation}
	ds^2= R_t^2(dt + \alpha dx)^2+R_x^2 dx^2 + R_y^2dy^2. 
\end{equation}
Permitting a parity twist on $T^3$, we can consider fermions satisfying  twisted boundary conditions 
\bea
&\psi(t, x+2\pi, y) = (-1)^{2a_x} \psi(t, x,  (-1)^{2b_x} y), \qquad \psi(t, x, y+2\pi) = (-1)^{2a_y} \psi(t, x, (-1)^{2b_y} y), & \nonumber\\ & \psi(t+2\pi, x - 2\pi\alpha, y) = (-1)^{2a_t} \psi(t, x,  (-1)^{2b_t} y),& \label{bc}
\eea
where each $a_i$ and $b_i$ can be $0$ or $\frac12$. Of the full modular group for $T^3$, only an $SL(2,\mathds Z)$ modular subgroup survives generated by
\begin{equation}
	U= \left(\begin{split}
	&0\,&1&\,&0\\
	\!\!\!-&1&0&&0\\
	&0&0&&1
	\end{split}\right),\qquad
	V= \left(\begin{split}
	&1&1&\,&0\\
	&0&1&&0\\
	&0&0&&1
	\end{split}\right).
\end{equation}
These transformations correspond to $S$ and $T$ of the $2$ torus with coordinates $(t,x)$. It is easy to see that under $V$, the parameter $\alpha\to\alpha+1$. As in the preceding discussion, only the $T$ transformation can be followed using our effective action reduction. 

We first review the modular properties of this theory following~\cite{Ryu:2015}.\footnote{Note that compared to~\cite{Ryu:2015}, our notation has $x$ and $t$ reversed to match the notation used in the rest of the paper.} We then discuss implications of these modular properties for the thermal partition function. This again leads to a prescription for calculating the anomaly via a correlation function. %We iterate that we are not concerned with constructing an effective theory which is fully $SL(2,\mathds Z)$ invariant and only concern ourselves with the $T$ transformation.

%%%%%%%%%%%%%%%%%%%%%%%%% sectors v2 %%%%%%%%%%%%%%%%%%%%%%%%

\subsection*{Sectors}

Since the background is a $T^3$, the theory has 64 sectors labeled by the choice of $(a_i, b_i)$ defined in~\C{bc}. 
In~\cite{Ryu:2015}, two choices of boundary condition in the $y$ direction were analyzed: $(a_y=0, b_y=0)$ and $(a_y=\frac12, b_y=0)$. These two choices are each separately invariant under the $SL(2,\Z)$ modular group. 
%are closed and do not mix with each other or any other sectors under $SL(2,\Z)$. 
The case$(a_y=\frac12, b_y=0)$ does not have an anomaly under a $T$ transformation, hence there will be no term in the effective action that is not invariant under $T$. We therefore specialize to the case $(a_y=0, b_y=0)$.

The remaining choices for the boundary conditions in $x$ and $t$ are divided into 4 subsectors with partition functions $\chi^i$ to $\chi^{iv}$. The superscript labels refer to the following boundary conditions, 
\bea \label{bcchi}
i: \left(b_t=0,\phantom{\frac12} b_x=0 \right), \quad ii: \left(b_t=0, b_x=\frac12\right), \nonumber \\ iii: \left(b_t=\frac12, b_x=0 \right), \quad iv: \left(b_t=\frac12, b_x=\frac12\right).
\eea
%satisfy boundary conditions $(G_t,G_x)=(\Gf^{2a_t},\Gf^{2a_x}),(\Gf^{2a_t},\P\Gf^{2a_x}),(\P\Gf^{2a_t},\Gf^{2a_x}),(\P\Gf^{2a_t},\P\Gf^{2a_x})$, where 
Each subsector is still characterized by a choice of $(a_t, a_x)$ so we denote the corresponding partition functions by $\chi^n_{[a_t,a_x]}$.
%As an example $\chi^{ii}_{[0,1/2]}$ is the partition function of the sector satisfying the boundary conditions $(1,\P\Gf)$ along the $t$ and $x$ directions respectively.
We are interested in the change of these partition functions under the $T^{-2}$ transformation. 

The partition functions in these sectors are given explicitly by a free-field computation~\cite{Ryu:2015}, 
\bea &\chi^{i}_{[a_t,a_x]}=A^R_{[a_t,a_x]}A^L_{[a_t,a_x]}
		\Theta^i_{[a_t,a_x]},\qquad
\chi^{ii}_{[a_t,a_x]}=A^R_{[a_t,a_x]}A^L_{[a_t,a_x-\frac12]}
		\Theta^{ii}_{[a_t,2a_x]}, &\nonumber\\
&\chi^{iii}_{[a_t,a_x]}=A^R_{[a_t,a_x]}A^L_{[a_t-\frac12,a_x]}
		\Theta^{iii}_{[2a_t,a_x]},\qquad
\chi^{iv}_{[a_t,a_x]}=A^R_{[a_t,a_x]}A^L_{[a_t-\frac12,a_x-\frac12]}
		\Theta^{iv}_{[2a_t,a_x-a_t]}, \label{parts} &
\eea
where the $\Theta^n$ are sums of massive $\Theta$ functions with masses given by the boundary condition for each sector; we have used their periodicity property:
\be \Theta_{[a+n,b+m]}=\Theta_{[a,b]},\;\;\;m,n\in\Z. \ee
The $A$ functions are the partition functions of a 2D chiral fermion. In terms of the 2D modular parameter  $\tau$, these partition functions have the following properties:
\begin{align}\label{eq:zeromodetrans}
&A^L_{[a,b]}(\tau)=(A^R_{[a,b]}(\tau))^*,\notag\\
& A^R_{[a,b]}(\tau)=A^R_{[a+1,b]}(\tau)\notag
	=e^{-2\pi i (a-1/2)}A^R_{[a,b+1]}(\tau),\\
&A^R_{[a,b]}(\tau+1)=e^{-\pi i (a^2-1/6)}A^R_{[a,b+a]}(\tau),\\
& A^R_{[a,b]}(-1/\tau)=e^{2\pi i (a-1/2)(b-1/2)}A^R_{[-b,a]}(\tau).\notag
\end{align}
We summarize the transformation properties of the $\Theta^n$ functions. Under $T^{-1}$,
\begin{align}\label{eq:unorientabletrans}
\Theta^i_{[a_t,a_x]}\to\;\Theta^i_{[a_t,a_t+a_x]},\;\;\;\notag
\Theta^{ii}_{[a_t,a_x]}\to\;\Theta^{ii}_{[a_t,a_x]},&\;\;\;
\Theta^{iii}_{[a_t,a_x]}\to\;\Theta^{iv}_{[a_t,a_x]},\;\;\;
\Theta^{iv}_{[a_t,a_x]}\to\;\Theta^{iii}_{[a_t,a_t+a_x]} \\
A^R_{[a_t,a_x]}\to\;e^{-\pi i (a_t^2-1/6)}A^R_{[a_t,a_t+a_x]}&,\;\;\;
A^L_{[a_t,a_x]}\to\;e^{\pi i (a_t^2-1/6)}A^L_{[a_t,a_t+a_x]}.
\end{align}
%and under $T^2=U_2^{-2}$:
%\begin{align}
%\Theta^i_{[a_x,a_t]}\to\;\Theta^i_{[a_x,a_t]},\;\;\;\notag
%\Theta^{ii}_{[a_x,a_t]}\to\;\Theta^{ii}_{[a_x,a_t]},&\;\;\;
%\Theta^{iii}_{[a_x,a_t]}\to\;\Theta^{iii}_{[a_x,a_x+a_t]},\;\;\;
%\Theta^{iv}_{[a_x,a_t]}\to\;\Theta^{iv}_{[a_x,a_x+a_t]}\\
%A^R_{[a_x,a_t]}\to\;e^{-2\pi i (a_x^2-1/6)}A^R_{[a_x,a_t+2a_x]}&,\;\;\;
%A^L_{[a_x,a_t]}\to\;e^{2\pi i (a_x^2-1/6)}A^L_{[a_x,a_t+ 2a_x]}
%\end{align}

Following the logic of our prior discussion, we study $T^{-2}$ which leaves the spin structure of each sector invariant. It is clear from~\eqref{eq:unorientabletrans}\ that any phase picked up under $T^{-2}$ must come from the transformation of the $A$ functions, i.e. from the zero modes. Looking at equation~\eqref{eq:zeromodetrans}, we see that the phase only depends on the first index of $A_{[a_t,a_x]}$, i.e. $a_t$. This implies that in~\C{parts}, the only partition functions that have a chance at generating a phase need a mismatch in the first index of $A^R$ compared with $A^L$. %sectors which do not have an imbalance in the first components of the right and left handed zero modes, do not pick up a phase. 
This is reminiscent of the requirement of level matching in 2-dimensional CFTs. %case where, even in the absence of a local anomaly, there can be a global anomaly if the boundary conditions of the left and right sectors are not the same. 
%Let us define $\tau_{2D} = \alpha + i{R_x\over R_t}$.  
%\[
%\chi^{iii}_{[a_t,a_x]}=A^R_{[a_t,a_x]}A^L_{[a_t-\frac12,a_x]}
%	\Theta^{iii}_{[0,a_x]}\longrightarrow
%	e^{2\pi i(-a_t+\frac14)}
%	A^R_{[a_t,a_x+2a_t]}A^L_{[a_t-\frac12,a_x+2a_t-1]}	\Theta^{iii}_{[0,a_x]}
%\]
%\begin{equation}
%	=e^{2\pi i(-a_t+\frac14)} \frac{A^R_{[a_t,a_x+2a_t]}A^L_{[a_t-\frac12,a_x+2a_t-1]}}{A^R_{[a_t,a_x]}A^L_{[a_t-\frac12,a_x]}}\chi^{iii}_{[a_x,a_t]}=\left\{
%	\begin{split}
%	&i \chi^{iii}_{[a_t,a_x]}&\;\;\;\;&a_t=0\\
%	-&i \chi^{iii}_{[a_t,a_x]}&\;\;\;\;&a_t=\frac12\\	
%	\end{split}
%	\right.
%\end{equation}
%and
%\[\chi^{iv}_{[a_t,a_x]}=A^R_{[a_t,a_x]}A^L_{[a_t-\frac12,a_x-\frac12]}
%\Theta^{iv}_{[0,a_x-a_t]}\longrightarrow
%e^{2\pi i(-a_t+\frac14)}
%A^R_{[a_t,a_x+2a_t]}A^L_{[a_t-\frac12,a_x+2a_t-\frac32]}
%\Theta^{iv}_{[0,a_x-a_t]}
%\]
%\begin{equation}
%=e^{2\pi i(-a_t+\frac14)} \frac{A^R_{[a_t,a_x+2a_t]}A^L_{[a_t-\frac12,a_x+2a_t-\frac32]}}{A^R_{[a_t,a_x]}A^L_{[a_t-\frac12,a_x-\frac12]}}\chi^{iv}_{[a_x,a_t]}=\left\{
%\begin{split}
%&i \chi^{iv}_{[a_t,a_x]}&\;\;\;\;&a_t=0\\
%-&i \chi^{iv}_{[a_t,a_x]}&\;\;\;\;&a_t=\frac12.\\	
%\end{split}
%\right.
%\end{equation}
The only sectors with an anomaly under $T^{-2}$ are therefore $\chi^{iii}$ and $\chi^{iv}$:
\begin{equation} \label{phasechi}
\chi^{iii,iv}_{[a_t,a_x]}\longrightarrow\left\{\begin{split}
&i \chi^{iii,iv}_{[a_t,a_x]}&\;\;\;\;&a_t=0\\
-&i \chi^{iii,iv}_{[a_t,a_x]}&\;\;\;\;&a_t=\frac12.	
\end{split}\right.
\end{equation}

%A quick analysis of the modular properties shows that $\chi^i$ and $\chi^{ii}$ are invarariant; however, $\chi^{iii}$ and $\chi^{iv}$ pick up a phase given by
%\begin{equation}
%\chi^{iii,iv}_{[a_t,a_x]}\longrightarrow\left\{\begin{split}
%&i \chi^{iii,iv}_{[a_t,a_x]}&\;\;\;\;&a_t=0\\
%-&i \chi^{iii,iv}_{[a_t,a_x]}&\;\;\;\;&a_t=1.	
%\end{split}\right.
%\end{equation}

%%%%%%%%%%%%%%%%%%%%%%%%%%%%%%%%%%%%%%%%%%%

\subsection*{Effective action}

The transformations of the partition function under the large diffeomorphism $T^{-2}$ given by \eqref{phasechi} must be matched by a local effective action in a gapped theory.  To proceed we promote the coefficient $\alpha$  to a field $a(x)$ and reduce the theory on the time circle. However, we immediately encounter a problem. There is no local 2-dimensional action that is both invariant under local diffeomorphisms and can reproduce the global anomaly. From the perspective of the EFT, this is simply because after a reduction along the time circle, the  2-dimensional theory is still not gapped. To see this, we note that the sectors that are not invariant under the action of $T^2$, i.e. $\chi^{iii}$ and $\chi^{iv}$, are the sectors that have a twisted boundary condition along the time direction, $b_t=1/2$.  These sectors have modes with zero momentum along the time direction regardless of the choice of $a_t$. As an example,  in the sector with $a_x=b_x=0$, it is  easy to check that the function $\psi(t,x,y)= \text{sign}(y)^{2a_t}\phi(y)$  satisfies the boundary conditions~\eqref{bc}\ for $\phi(y)$ even in $y$. Therefore, even if we impose twisted anti-periodic boundary conditions, the gap that is generated is of  order  $1/R_y$ as opposed to $1/R_t$.

In order to derive a gapped effective action, we must therefore work at scales $\lambda$ which not only satisfy  $\lambda>>R_t$ but also  $\lambda>>R_y$. This would imply that we are effectively looking at a 1-dimensional action. Indeed, in one dimension, we can write down an effective action which correctly reproduces the phase factors appearing in~\C{phasechi}:
\begin{equation}\label{3danomaly}
S=(-1)^{2a_t}\frac{i\pi}{4}\int a. 
\end{equation}
As an immediate consequence of~\C{3danomaly}, we see that it is possible to derive the anomaly under the $T$ transformation from a stress-tensor $1$-point function. 

%is that after reducing on $x$ to 2D, the system still has a zero mode. The boundary condition we are choosing is twisted along the $x$ direction so that even if we pick $a_x=1/2$ there is a zero mode if the fermionic field is odd in $y\to-y$. This implies that from the 1D perspective, the mass of the lowest lying state is $\sim 1/L_y$ as opposed to $1/L_x$. Hence from the 2D $(t,y)$ perspective, the system is still gapless and we don't expect a local partition function to describe the full dynamics. The same conclusions as  in the 2-dimensional case can be derived here. For example, the anomaly factor can again be derived using a stress expectation value.

%%%%%%%%%%%%%%%%%%%%%%%%%%%%%%%%%%%%%%%%%%%%%%%%%%%%%%%%%%%%%%%%%
\subsection{Weyl fermions in 4D and higher}
\label{sec:4D}
%%%%%%%%%%%%%%%%%%%%%%%%%%%%%%%%%%%%%%%%%%%%%%%%%%%%%%%%%%%%%%%%%

In this section we look at the global anomaly properties of Weyl fermions in 4D in the presence of a background $U(1)$ gauge field $A_\mu$. In \cite{Golkar:2012}, one of us conjectured that the factor of ${1\over 12}$ in the coefficient of the chiral vortical effect is the sign of a global anomaly. Here we prove that this is in fact the case and show that global anomaly considerations fix this coefficient mod 2.

Similar to our other examples, we wish to derive a local effective action by considering the system on a thermal circle. At large temperature, the theory is effectively 3-dimensional and gapped. In order to have non-trivial but simple global transformations, we define the system on $\mathcal M = T^2 \times X_2$ where $X_2$ is a compact 2-dimensional manifold with metric
\be ds^2 = \left(dt+ a_i(x) dx^i \right)^2 + ds_X^2, \ee
and  $ds_X^2$ is the line element on $X_2$. We treat the $T^2$ in the same manner as the 2-dimensional case discussed above. Again there is an $SL(2,\mathds Z)$ associated to the modular group of $T^2$, and again we restrict to the $T$ transformation which preserves the effective field theory limit.

\subsubsection*{Global gravitational anomaly} 

We again consider the change of the partition function under the $T$ transformation: $t\to t+x$, $a_x \to a_x +{\beta \over L}$. This time, however, we do so in the presence of some $U(1)$ flux on the compact manifold $X_2$. It is straightforward to compute the anomaly using the $\eta$-invariant and its relationship to a 6-dimensional index theorem, similar to the 2-dimensional case presented in appendix~\ref{app:2D_mappingtorus}. However, here we will use a simpler approach.% that can grant us more insight. 

The crucial observation is that our global transformation leaves $X_2$ invariant, hence we can reduce the theory to $T^2$. Since massive Dirac fermions do not contribute to the global anomaly, we only need to consider any chiral zero modes that arise from this reduction. These are given by the 2D Atiyah-Singer index theorem for the manifold $X_2$:
\be \nu=\nu_- - \nu_+ = N_\phi= \frac1{2\pi}\int_{X_2} F. \ee
Reducing to $2$ dimensions, we find a net $\nu$ chiral zero modes. Hence the phase produced by a $T$ transformation is equal to the phase of a single chiral mode in 2D, given by $\pi\over 24$, multiplied by the number of zero modes $\nu$. We conclude that 
\be S_{eff}\rightarrow S_{eff}+ \frac{i\pi}{24}N_\phi . \ee
 %have derived that\footnote{Note that similar to section 2, we either have to pick a spin structure which is invariant under $T$ or look at $T^2$. }  
This needs to be reproduced in the effective action by a local term. We therefore list all the local terms that are local functionals of the background fields and have 3D Lorentz invariance. Since we assume there are no perturbative anomalies, or that any perturbative anomalies have been matched by other terms, %\footnote{One can equivalently say that perturbative anomalies have been matched via other terms hence the remainder should be invariant under perturbative transformations.}, 
these local terms have to be invariant under perturbative diffeomorphisms and gauge transformations. There is only one such term: $\int a \wedge d \mathcal A$, where we have defined $\mathcal A_i = A_i - A_0 \, a_i$ so that $\mathcal A_i$ and $a_i$ are independent $U(1)$ connections~\cite{Banerjee:2012}. Matching the coefficient of this term with the global anomaly gives, 
\begin{equation}
\label{eq:CVE}
S_{eff}=\frac{i}{12\times 4\pi}\int a\wedge d\mathcal A,
\end{equation}
which is the calculated effective action for the chiral vorticity effect, a potentially physically measurable phenomenon where a chiral current is induced in the direction of vorticity in a fluid. 

\subsubsection*{Global gauge anomaly}

We might ask what~\C{eq:CVE}\ implies for situations where we turn on a graviphoton flux compatible with the fermion spin structure. Whether this is possible is subtle because the graviphoton flux changes the topology of the $4$-dimensional space.\footnote{As an example of this subtlety, consider the case of $\mathds R^3\times S^1$. Topologically non-trivial choices for the graviphoton correspond to replacing the boundary at infinity, $S^1\times S^2$, with a Hopf fibration of $S^1$ over $S^2$. The non-trivial topological choices correspond to a choice of magnetic charge captured by the first Chern class of the graviphoton gauge bundle. The Taub-NUT manifold is an example of this type. The non-trivial topology at infinity has the effect of trivializing $\pi_1$ for the space, allowing the circle to unwind.} Let us assume this is possible. 

Equation~\eqref{eq:CVE}\ then implies that $S_{eff}$ is also not invariant under large gauge transformations in the presence of graviphoton flux. This might seem counter-intuitive because~\C{eq:CVE}\ is a function of $d\mathcal A$, which is gauge invariant. However, we point out that when there is a flux in $a$, the fields need to be defined in patches with transition functions in the overlap regions and these transition functions are not gauge invariant~\cite{Polychronakhos}. A careful computation reveals, 
\be \mathcal A\to \mathcal A + 2\pi, \quad S_{eff} \to S_{eff} + \frac{i}{24}\int da.\ee
Note that this is the same result we would find by naively integrating the Chern-Simons term by parts, ignoring boundary and transition terms.

Since it is not possible to write down a term in the effective action which would induce a global gravitational anomaly without a possible global gauge anomaly and vice-versa, we conclude that the presence of one requires the presence of the other. This is the reciprocity property of mixed anomalies which can also be seen, albeit less directly, from the computation of the $\eta$ invariant for the two transformations; see appendix~\ref{app:reciprocity}. 

\subsubsection*{Higher dimensions}

From the 4-dimensional derivation, it is easy to see how to generalize to higher dimensions. Consider a Weyl fermion on a $d$-dimensional manifold $\mathcal M_d=T^2\times X_{d-2}$. We can reduce the theory on $X_{d-2}$ to get an effective 2-dimensional chiral theory on $T^2$. Again the number of  chiral modes is determined by index theory on  $X_{d-2}$. This reduction suggests that the anomalous terms in the thermal effective action take the form, 
%part of the thermal partition function for these theories at low energies would be:
\begin{equation}
S_{eff}=\int_{S^1 \times X_{d-2}} \hspace{-10pt} a \wedge \widehat{A}(X_{d-2})\wedge \text{ch}(V),
\end{equation}
where $a$ is the graviphoton as before and $V$ is the gauge bundle, which was $U(1)$ in our prior discussion. For example in 6 dimensions we would find, 
\be 
S_{6D}=\int a\wedge \left\{\frac1{24\times 8\pi^2 } \text{Tr} (R\wedge R) - \frac1{8\pi^2}{\rm Tr} (F\wedge F) \right\},
\ee
with analogous expressions in higher dimension.

\section*{Acknowledgments}
It is our pleasure to thank  Jeffrey Harvey, Sungjay Lee, Matthew Roberts, Dam T. Son and David Tong for insightful discussions.  This work is supported in part by DOE grant DE-FG02-13ER41958. S.~G. is supported in part by NSF MRSEC grant DMR-1420709 and European Research Council under the European Union’s Seventh Framework Programme (FP7/2007-2013), ERC grant agreement STG 279943, ``Strongly Coupled Systems''. S.~S. is supported in part by NSF Grant No.~PHY-1316960.

\newpage
\appendix

\section{Global gravitational anomaly in 2D}
\label{app:2D_mappingtorus}

The modular properties of Majorana-Weyl fermions on 2-dimensional tori are well studied. Here we will recount the derivation of the change of the action under a $T$ transformation from the perspective of the $\eta$ invariant calculation for the case of periodic boundary condition along the time direction.\footnote{This choice is dictated to us by topological considerations when constructing the 4-dimensional manifold.} We start with the metric,
\be ds^2=(dt+a(x) dx)^2 + dx^2,\ee
with periodicities  $t \sim t + \beta,\; x\sim x+L$. We note that all curvatures vanish in this background and hence there is no perturbative anomaly. The large diffeomorphism of interest takes $a(x) \to a(x)+{\beta \over L}$. We therefore construct the mapping torus $\Sigma$ with metric,
\be ds^2=dy^2+\left(dt+\left[ a(x) + {\beta y \over L} \right] dx\right)^2 + dx^2,\ee
where the $y$ coordinate interpolates between the original torus and the torus with shifted modular parameter. Finally, we have the identification 
\be (t,x,y) \sim (t- {\beta x \over L},x,y+1),\ee
of the torus at $y=0$ with its image under the large diffeomorphism at $y=1$. 
%To see this note that the metric with $t'$ and $x'$ at $y=1$ is $ds^2=(dt'+(a+\beta/L)dx)^2$ and the metric at $y=0$ with  $t$ and $x$ can be related by a large diff: $x'=x$, $t' = t-\beta x/L$. Hence, $t$ the coordinate at $y=0$ must be identified with $t'=t-\beta x/L$, the coordinate at $y=1$. 
We need to calculate the $\eta$ invariant on this manifold. This can been done directly via the computation of the eigenvalues of the Dirac operator \cite{Solvmanifold} but here we will use the APS index theorem.

%We define the tetrad:
%\begin{equation}
%{e^a}_\mu=\left(
%\begin{split}
%&1\;\;\;  &  a(&x)+y\beta/L\;\;   &   0\\
%&0  &  &\;\;\;\;\;1    &   0\\
%&0  &  & \;\;\;\;\;0    &   1\\
%\end{split}
%\right),
%\end{equation}
%and the non-zero components of the spin connection ${\omega^{ab}}_\mu$ are:
%\[{\omega^{ty}}_x=\frac{\beta}{2L},\;{\omega^{xy}}_t={\omega^{xt}}_y=\frac{\beta}{2L\sqrt{h(x)}},\;
%{\omega^{xy}}_x=\frac{\beta\big(y\beta+La(x)\big)}{2L^2 \sqrt{h(x)}}. \]
%For simplicity, we take $a(x)=a$ and $h=L=1$. We then have:
%\[{\omega^{ty}}_x={\omega^{xt}}_y={\omega^{xy}}_t=\frac{\beta}{2},\;\;\; {\omega^{xy}}_x=\frac12 \beta (a+y\beta) . \]
%The 3D dirac eigenvalue problem is:
%\[D\psi = i\gamma^\mu  (\partial_\mu +\tfrac12 \Sigma^{ab}{\omega^{ab}}_\mu)\psi = E\psi, \]
%where $\Sigma^{ab}=\frac14[\gamma^a,\gamma^b]$.
%
%We take the following ansatz:
%\[\psi(t,x,y)=e^{i(\omega t + k x+\omega y)}\psi(y),\]
%such that $e^{-i\omega\beta x/L}\psi(y+1)=\psi(y)$.
%\note{Direct calculation is hard}

To proceed, we must find a 4-dimensional manifold $X$ with the mapping torus $\Sigma$ as its boundary, $\partial X = \Sigma$. Since $\Sigma$ is topologically a 3-torus, $X$ can be constructed by filling in any of the circles of $\Sigma$. However, it is not possible to fill in either the $x$ or the $y$ circles because neither choice is consistent with the boundary conditions imposed at $y=0$ and $y=1$. We therefore proceed by filling in the $t$ circle,\footnote{Note that this restricts the boundary conditions imposed on the fermion along the time circle since the spin structure of the manifold must be extendible to $X$.}
\begin{equation}\label{eq:4d_metric}
ds^2=dr^2 +dy^2+f(r)^2 \left(dt^2+  \left[a(x)+\frac{\beta y}{L} \right]dx\right)^2+dx^2, 
\end{equation}
where $r$ ranges from $0$ to $1$, and $f(r)$ is a radial function that must be linear at $r=0$ to avoid a conical singularity. In order to avoid extraneous factors,  we also assume that the periodicity of the $t$ direction is $2\pi$.

The  APS index theorem for the spin complex of a 4-dimensional manifold reads,
\begin{equation}
\label{eq:APS_4D}
{\rm Ind}(\, \Dslash) = \frac1{24\times 8\pi^2} \int_X \text{Tr }(R\wedge R)-\frac1{24\times 8\pi^2} \int_\Sigma \text{Tr }(\theta \wedge R) -\frac\eta2, 
\end{equation}
where ${\theta^a}_b$ is the second fundamental form given by the difference of the spin connection ${\omega^a}_b$ derived from the metric \eqref{eq:4d_metric}, and the spin connection coming from the product metric at the boundary ${(\omega_0)^a}_b$. Since $\eta$ is only defined mod 2, we can calculate the remaining terms on the right hand side of~\C{eq:APS_4D}\ and determine $\eta$ by demanding integrality of the index on a compact space. We see via direct computation on this background that:
\begin{equation}
\label{eq:APS_4D_compute}
{\rm Ind}(\, \Dslash)=  \frac{1}{24}- {\pi^2 \over 12} {f(1)^4 \over L^2}-\frac\eta2.
\end{equation}
We see that there is a contribution to the index which is not purely topological and depends on various size factors; namely, the size of the $x$-circle as well as volume of the 2-torus. However, we notice that this is exactly the contribution of the 3-dimensional gravitational Chern-Simons term:
\[\frac1{24\times 8\pi^2} \int_{\Sigma} \left( \omega_{(3)}\, d\,\omega_{(3)} +\frac23\,{\omega_{(3)}}^3 \right)={\pi^2 \over 12} {f(1)^4 \over L^2},\]
where $\omega_{(3)}$ is the intrinsic spin connection on the 3-dimensional manifold $\Sigma$.  We therefore conclude that:
\begin{equation}
\label{eq:eta_answer}
\eta=\frac1{12} + \frac1{96\,\pi^2} \int_\Sigma \left(\omega_{(3)}\, d\,\omega_{(3)}+\frac23\,{\omega_{(3)}}^3 \right). 
\end{equation}
As expected, the difference between $\eta$ and the Chern-Simons term is a topological invariant and gives the modular transformation of the 2D Weyl fermion. 

\section{Reciprocity and global anomalies}\label{app:reciprocity}
In the example discussed in section \ref{sec:4D}, we saw that a study of the effective action implied that a global  gravitational anomaly in background magnetic flux requires a reciprocal global gauge anomaly in gravitational flux. In this section, we describe how this happens in the computation of global anomalies using the  $\eta$ invariant.

We work with the manifold discussed in section~\ref{sec:4D}, where $\mathcal M = T^2 \times X_2$. Here, for simplicity we take the two-dimensional compact manifold $X_2$ to also be a torus. As in section~\ref{sec:4D}, we thread one quantum of magnetic flux through $X_2$. We take the periodicities of all circles to be $2\pi$. The background fields are, 
\begin{equation}
\label{eq:4D_Mixed_BGs}
ds^2 = \left(dt+a(x)dx\right)^2 + dx^2  +dy^2 +dz^2, \qquad A_z= y,
\end{equation}
with other components of the gauge field equal to zero. We consider the large diffeomorphism $t\to t+x$. For this transformation, we construct the mapping torus $\Sigma$. % which interpolates between the manifold and its diffeomorphism in the higher dimension that we parametrize as $r$ which runs from $0$ to $1$. 
We denote the interpolating coordinate by $r$ which runs from $0$ to $1$. The metric on $\Sigma$ and the background gauge-field take the form,
\begin{equation}
\label{eq:5D_Mapping}
ds^2 = dr^2 + \left(dt+(a(x)+r)dx\right)^2 + dx^2  +dy^2 +dz^2, \qquad A_z= y.
\end{equation}
Now the crucial point in the argument is that the mapping torus described in \eqref{eq:5D_Mapping} can also be considered as a mapping torus for a large gauge transformation $A_z \to A_z + 2\pi$ on a background:
\begin{equation}
\label{eq:4D_Mixed_BGs_alternate}
ds^2 = dr^2 + \left(dt+(a(x)+r)dx\right)^2 + dx^2+  dz^2, \qquad A_z= 0.
\end{equation}
where now we have a background gravitational flux stored in the gauge field $a$ from the perspective of the dimensionally-reduced theory. The interpolating coordinate is now $y$. Therefore, in this case, the calculation of the global gravitational anomaly in a gauge magnetic flux is precisely the same as the calculation of a global gauge anomaly in a gravitational flux. 

Note that not every mapping torus associated to a mixed global anomaly can be interpreted as the mapping torus for two different transformations in this manner. This is possible in this case because the space is toroidal and the gauge group is $U(1)$. However, if we take $X_2$ to be $S^2$, we would no longer be able to interpret $\Sigma$ as a mapping torus for a large gauge transformation.

\addcontentsline{toc}{section}{Bibliography}
\bibliographystyle{JHEP}
\bibliography{GBA}

%%%%%%%%%%%%%%%%%%%%%%%%%%%%%%%%%%%%%%%%%%%%%%%%%%%%%%%%%%%%%%%%%
\end{document}